\documentclass[showpacs,preprintnumbers,aps,prb]{revtex4}
\usepackage[pdftex]{graphicx,color}
\usepackage{dcolumn}
\usepackage{bm}
\usepackage{longtable}
\usepackage{amsmath,amssymb}

\newcommand{\pauli}[1]{\ensuremath{\sigma^{#1}}}

\usepackage{amssymb,amsmath}
\usepackage{hyperref}
\usepackage{pgffor}
\usepackage{bbm}

\newcommand{\tr}{\ensuremath{\mathrm{tr}\,}}					

\newcommand{\ba}{\begin{eqnarray}}
\newcommand{\ea}{\end{eqnarray}}
\newcommand{\cnot}{\ensuremath{\mathrel{\text{CNOT}}}}				
\newcommand{\Ham}{\ensuremath{\mathcal{H}}}					
\newcommand{\id}{\ensuremath{\mathbbm{1}}}					
\newcommand{\subsc}[1]{_{\text{#1}}}						

\newcommand{\ket}[1]{\ensuremath{\left|\, #1\right>}}

\makeatletter
\newcommand{\fmslash}[2][0mu]{%
  \mathchoice
    {\fmsl@sh\displaystyle{#1}{#2}}%
    {\fmsl@sh\textstyle{#1}{#2}}%
    {\fmsl@sh\scriptstyle{#1}{#2}}%
    {\fmsl@sh\scriptscriptstyle{#1}{#2}}}
\newcommand{\fmsl@sh}[3]{%
  \m@th\ooalign{$\hfil#1\mkern#2/\hfil$\crcr$#1#3$}}
\makeatother

\begin{document}
\title{Representation of excited states and topological order of the toric code in MERA}
\author{Johannes M. Oberreuter}
\email{johannes.oberreuter@theorie.physik.uni-goettingen.de}
\affiliation{Technische Universit\"{a}t M\"{u}nchen, Walter Schottky Institut, Am Coulombwall 4, 85748 Garching, Germany}
\affiliation{Georg-August-Universit\"{a}t G\"{o}ttingen, Institute for Theoretical Physics, Friedrich-Hund-Platz 1, 37077 G\"{o}ttingen, Germany}
\author{Stefan Kehrein}
\affiliation{Georg-August-Universit\"{a}t G\"{o}ttingen, Institute for Theoretical Physics, Friedrich-Hund-Platz 1, 37077 G\"{o}ttingen, Germany}
\begin{abstract}
The holographic duality relates a field theory to a theory of (quantum) gravity in one dimension more. 
The extra dimension represents the scale of the RG transformation in the field theory. 
It has been conjectured that the tensor networks which arise during the real space renormalization procedure like the multi-scale entanglement renormalization ansatz (MERA) are a discretized version of
the background of the gravity theory.
We strive to contribute to make this conjecture testable by considering an explicit and tractable example, namely the dual network of the toric code, for which MERA can be performed analytically. 
We examine how this construction can be extended to include excited states. Furthermore, we show how to calculate topological entanglement entropy from the geometry of MERA. This method is expected to generalize to systems with  generic entanglement structure.
\end{abstract}

\maketitle

\section{Introduction}

Condensed matter systems at strong coupling are found in a large variety of circumstances, for instance in materials that exhibit high-temperature superconductivity \cite{Keimer2015}, however, they are often hard to examine. Often such systems at strong coupling show long-range entanglement or absence of quasiparticles. It is desirable to explore novel methods to solve such systems apart from numerics. 

Very often, the use of dualities makes a problem easier to solve mapping it to a more suitable model. The most notable such duality in condensed matter physics is the Jordan-Wigner transformation, which maps a spin model to free fermions. The gauge-gravity duality is a comparably new such technique, which can help with the treatment of strongly coupled systems near a critical point. It has been noted \cite{Maldacena:1997re,Witten:1998qj,Gubser:1998bc} that a specific conformal gauge field theory, namely $\mathcal{N}=4$ Super-Yang-Mills theory with an $SU(N)$ gauge group is dual to a geometric theory of gravity, in particular to type IIB supergravity, which is a low energy effective description of string theory, on a space-time with a negative cosmological constant. Since such a maximally symmetric vacuum solution of the Einstein equations of general relativity with negative curvature are called \emph{Anti-de Sitter space} (AdS)  this duality has been named \emph{AdS/CFT correspondence}. It is useful because it relates strongly coupled field theories to weakly coupled classical gravity and vice versa. 
Notable examples of condensed matter systems at strong coupling fixed points which have been treated with the use of the AdS/CFT correspondence include the superfluid-insulator transition of ultra-cold bosonic atoms on an optical lattice \cite{Witczak-Krempa2014} or non-quasiparticle transport in non-Fermi liquids \cite{Liu:2009dm,Cubrovic:2009ye}.

Usually, the correspondence is taken on a restricted coordinate set of AdS space, the so-called Poincar\'{e} patch. Since in these coordinates, the boundary of AdS for the radial coordinate taken to infinity is Minkowski space in one dimension less, one usually talks about the field theory to live on the flat boundary, while the AdS-space is referred to as the gravitational bulk. Since the lower dimensional theory still contains the information about the higher dimensional system, this duality has been nicknamed \emph{holography} due to the resemblance with the optical phenomenon \cite{tHooft1993,Susskind1995}. In the original and still best understood version, the correspondence relates five-dimensional AdS gravity to four-dimensional gauge theory on Minkowski space. From the point of view of the field theory, the radial direction of AdS space is an extra emergent coordinate. This direction can be interpreted as a renormalization group energy scale, which takes us from the UV of the field theory at the boundary of AdS to its IR deep in the bulk by integrating out high energy modes (cf. \cite{deBoer:1999xf,Faulkner2011,Heemskerk:2010hk}).

It needs to be noted, though, that this useful correspondence between gravity and many-body theories is strictly speaking only exact under a rather limiting set of conditions \cite{Maldacena2011}. Since the isometries of AdS space replicate the scale invariant structure, in general, the field theory in such a correspondence needs to be conformal. To ensure this in the known cases, the field theory has to have a high amount of supersymmetry.  In the canonical case, this is the maximal amount of four supersymmetry charges ($\mathcal{N}=4$). Finally, only gauge theories allow for the (large-N) scaling limit which corresponds to classical gravity in the bulk. These restrictions are usually not readily found in generic condensed matter systems, which are for instance defined on a lattice away from a critical point and usually do not feature any supersymmetry. This restricts the use of an otherwise nifty method.

The objective of our research program is to extend the scope of this correspondence and to devise an alternative, more general formulation of the duality between a geometric theory and a quantum many body theory. The important clue is the interpretation of the extra emergent direction as the RG-scale. In this regard, a real space RG procedure with a hyperbolic structure like AdS would be a candidate of a generalized geometry/condensed matter duality.
It has been suggested \cite{Swingle2012a,Swingle2012} that the multiscale entanglement renormalization ansatz (MERA) \cite{Vidal2007} can be interpreted as a way to represent a duality between a (geometric) renormalization circuit network and the quantum theory it is renormalizing on its UV lattice. 
The essential point of this real space renormalization procedure is that during each RG step a local unitary basis transformation reduces local entanglement between neighboring sites before coarse graining the system with isometries. Thus, the growth of the entanglement entropy is alleviated. This routine gives rise to a network of tensors, which forms a hyperbolic structure and resembles the AdS geometry. The big advantage of such a picture would be that the MERA can in principle be constructed for almost any many-body system. The question remains, in which way this is a correspondence. 
To test this idea it is very desirable to examine explicit examples, for which the tensor network can be constructed analytically. 

There is, indeed, an interesting system, namely Kitaev's toric code, for which the real space renormalization procedure is well-understood \cite{Aguado2008}. Using the algorithm for adding qubits to and removing them from the toric code, the form of the disentanglers and coarse grainers can be constructed explicitly using elementary CNOT operations.
Apart from this advantage, the model is interesting because it shows topological order, noticeable as its ground state is fourfold degenerate and the system has four sectors of excitations. So far, it is not clear how to incorporate topological order in a holographic setting. We are investigating this question and suggest a geometric picture for it.
Furthermore, we generalize the MERA of the toric code to also encode excited states and we find the same structure of the network as for ground states, which is a property specific to the toric code.
The geometric construction of the topological entanglement entropy, however, is generic and does not suffer from the peculiarities of the model, which rather allow to work it out explicitly in this system.

This paper is organized as follows: In section \ref{sec:tc}, we review the basic definition and properties of the toric code including its excited states. We also explain how the algorithm for adding qubits to or removing them from a toric code works for excited states, as well. In section \ref{sec:excited}, we use this to extend the MERA of the toric code to also represent excitations. In section \ref{sec:tee}, we explain how to calculate the entanglement entropy from MERA as a discrete geometry. In particular, we give an interpretation to the topological entanglement entropy in terms of MERA.

\section{Short introduction into the spectrum of the toric code}
\label{sec:tc}

\newcommand{\checkp}{\ensuremath{\Box}}
\newcommand{\checks}{\ensuremath{+}}
\newcommand{\stabil}{\ensuremath{\xi}}

\subsection{Definition}

\begin{figure}
\centering
\resizebox{8cm}{!}{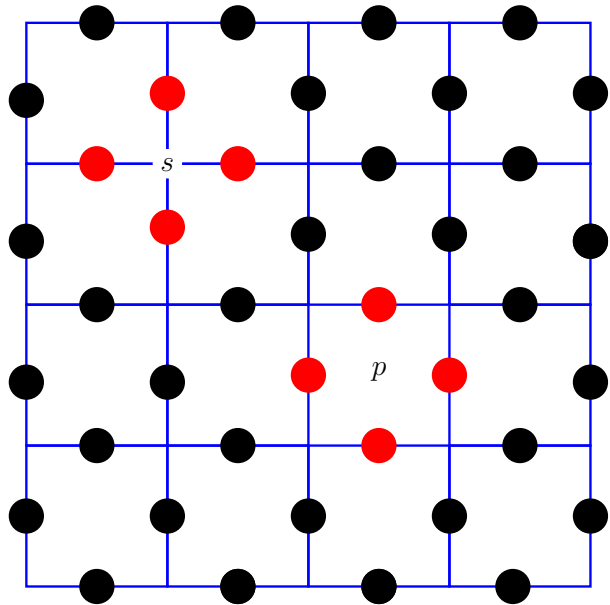}
\caption{Although the physical degrees of freedom of the toric code are qubits located on the links of a square lattice (dots), the they can better be organized in terms of star $s$ and plaquette $p$ operators, to which the stabilizer conditions \eqref{eq:stabilizer} are applied. (Color online)}
\label{fig:tc_dof}
\end{figure}

The toric code \cite{Kitaev1997,Kitaev2003} is a spin model defined on a square lattice, where the individual spins/qubits are located on the links of the lattice. It is defined in terms of the so-called check operators
\begin{equation}
 \checkp_p = \otimes_{l \in p} \pauli{z}_l \;, \quad \checks_s = \otimes_{l \in s} \pauli{x}_l \;, 
 \label{eq:stabilizer}
\end{equation}
where the operators act on all the links $l$ in a plaquette $p$ or a star $s$, respectively (cf. fig. \ref{fig:tc_dof}). A simultaneous eigenstate $\ket{\stabil}$ of all check operators, for which $\checkp \ket{\stabil} = \ket{\stabil}$ or $\checks \ket{\stabil} = \ket{\stabil}$, is said to be stabilized. The model is defined in terms of the check operators via the Hamiltonian
\begin{equation}
\Ham = - \lambda_p \sum_p \checkp_p - \lambda_s \sum_s \checks_s \;,
\label{eq:Ham_tc}
\end{equation}
where we choose $\lambda_p=\lambda_s=1$.
The ground states of the Hamiltonian are stabilized on all plaquettes and stars and therefore called stabilizer states. 

The ground state of the toric code is four-fold degenerate. The ground states can be classified according to the eigenvalue of a closed chain of $\pauli{z}$ operators applied to adjacent links spanning a non-trivial cycle of the torus. All eigenstates with the same eigenvalue are part of one ground state. Since there are two cycles on a torus with two eigenvalues $\pm 1$ each, the degeneracy is four-fold. Note that all the chain operators along the same cycle are equivalent in the sense that they can be deformed into each other by applying plaquette operators. We can switch between different ground states by applying a chain of $\pauli{x}$ operators on every qubit around any full cycle of the torus. The resulting ground state contains all the stabilizer states with the new $\pauli{z}$ chain eigenvalue.

\subsection{Excited states}

There are two kinds of excitations which can live in each of the four ground-state manifolds, both of which violate the stabilizer conditions:
\begin{itemize}
 \item Acting with $\pauli{x}$ on any link creates a flux-(anti-)flux pair on the adjacent plaquettes, violating their plaquette stabilizer
 \item Acting with $\pauli{z}$ on any link creates a charge-(anti-)charge pair on the adjacent sites, violating their star stabilizer
\end{itemize}
Hence, one excitation consists of two ``particles'', living on two different plaquettes or stars, respectively.
Since the Pauli-matrices square to the identity $(\pauli{x})^2 = (\pauli{z})^2 = \id$, the excitations are their own anti-excitation and two particles of the same kind annihilate. Correspondingly, creating another excitation pair next on a link adjacent to one of the particles moves this excitation one place further. This means that the excitation gets ``stretched'' and therefore non-local.

The energy spectrum of the excitations can be calculated realizing that every excitation turns two stabilizers of each kind from a value $+1$ to $-1$, yielding an energy difference of $2$ per particle. 
In particular
\begin{eqnarray}
 \Ham \ket{X_i} &=&  \left(- \sum_{s} \checks_s - \sum_p \checkp_p \right) \pauli{x}_i \ket{\stabil} \\
 &=& - \pauli{x}_i \left( \sum_{i \notin s} \checks_s + \sum_{i\notin p} \checkp_p  \right) \ket{\stabil} \nonumber \\
 && - \pauli{x}_1\pauli{x}_2\pauli{x}_3\pauli{x}_i \pauli{x}_i \ket{\stabil} - \pauli{x}_4\pauli{x}_5\pauli{x}_6\pauli{x}_i \pauli{x}_i \ket{\stabil} \\
 && - \pauli{z}_1\pauli{z}_2\pauli{z}_3\pauli{z}_i \pauli{x}_i \ket{\stabil} - \pauli{z}_4\pauli{z}_5\pauli{z}_6\pauli{z}_i \pauli{x}_i \ket{\stabil} \\
 &=& \left( - ( N_p + N_s - 4 ) +2-2 \right) \pauli{x}_i \ket{\stabil}\\
 &=& \left(- N_p - N_s +4 \right) \ket{X_i}
\end{eqnarray}
and analogously for a $\ket{Z_i}$ excitation.
So we have
\begin{itemize}
 \item Flux: $\ket{X} = \pauli{x} \ket{\stabil}$ with energy $H \ket{X} = -(N_s+(N_p-2)-2) \ket{X} = -(N_s + N_p -4) \ket{X}$
 \item Charge: $\ket{Z} = \pauli{z} \ket{\stabil}$ with energy $H \ket{Z} = -((N_s-2)+N_p-2) \ket{Z} = -(N_s + N_p -4) \ket{Z}$.
\end{itemize}

For the isotropic toric code, the excited states are degenerate beyond the topological degeneracy. We will focus on the degeneracy within one ground state manifold first. A state of the toric code with $n$ excitations has an energy of
\begin{equation}
E_n = -(N_s + N_p - 4n) 
\end{equation}
which can be distributed over charge or flux excitations. Since all excitations can also be delocalized, an energy state is degenerate with the degree of the number of possibilities to distribute each excitation over the lattice. Essentially, for every excitation we choose two sites and plaquettes, respectively, whose stabilizer is violated and where the particles are situated. Hence, the degeneracy is
\begin{equation}
 \mathrm{deg}_n = \binom{N_p}{2n} + \binom{N_p}{2(n-1)} \binom{N_s}{2} \dots + \binom{N_s}{2n} \;.
 \label{eq:deg_excitation}
\end{equation}
Note that the degeneracy decreases with increasing number of excitations, after half of the toric code lattice is filled. The maximum energy
\begin{equation}
 E_{\text{max}} = - \left( N_s + N_p - 4(\frac{N_s}{2}+\frac{N_p}{2}) \right) = N_s + N_p
\end{equation}
is reached when all plaquettes and all sites are filled with one excitation and is not degenerate
\begin{equation}
 \mathrm{deg}_{\text{max}} = \binom{N_p}{2 \frac{N_p}{2}} \binom{N_s}{2 \frac{N_s}{2}} = 1 \;.
\end{equation}

\subsection{Adding qubits to and removing them from the toric code}
\label{subsec:addremove}

A very useful feature of the toric code is that additional qubits and thus links can be added to and removed from it. This feature is exploited to construct the tensors of the MERA (i.e. disentanglers and coarse grainers) for the toric code explicitly. This process changes the lattice of the toric code and it has been realized in \cite{Dennis2002} how simultaneous $\cnot$ operations involving the new qubit change the stabilizer conditions accordingly such that a new plaquette or star is added as we explain in the following. The $\cnot$ gates are logical operations which take a so-called control- and a target bit as input. It negates the target bit if and only if the control bit is true, which is itself not changed by the operation. We adopt the usual convention that $\ket{\uparrow} = \ket{0} = \pauli{z} \ket{0}$ and $\ket{\downarrow} = \ket{1} = -\pauli{z} \ket{1}$. Adding and removing a link to and from the toric code are the same process due to the fact that the $\cnot$ gate squares to $\cnot^2 = \id$.

\begin{figure}
\centering
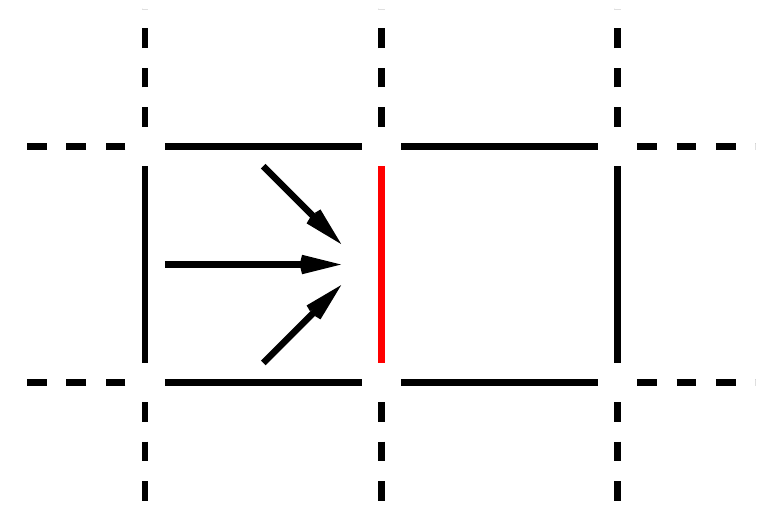
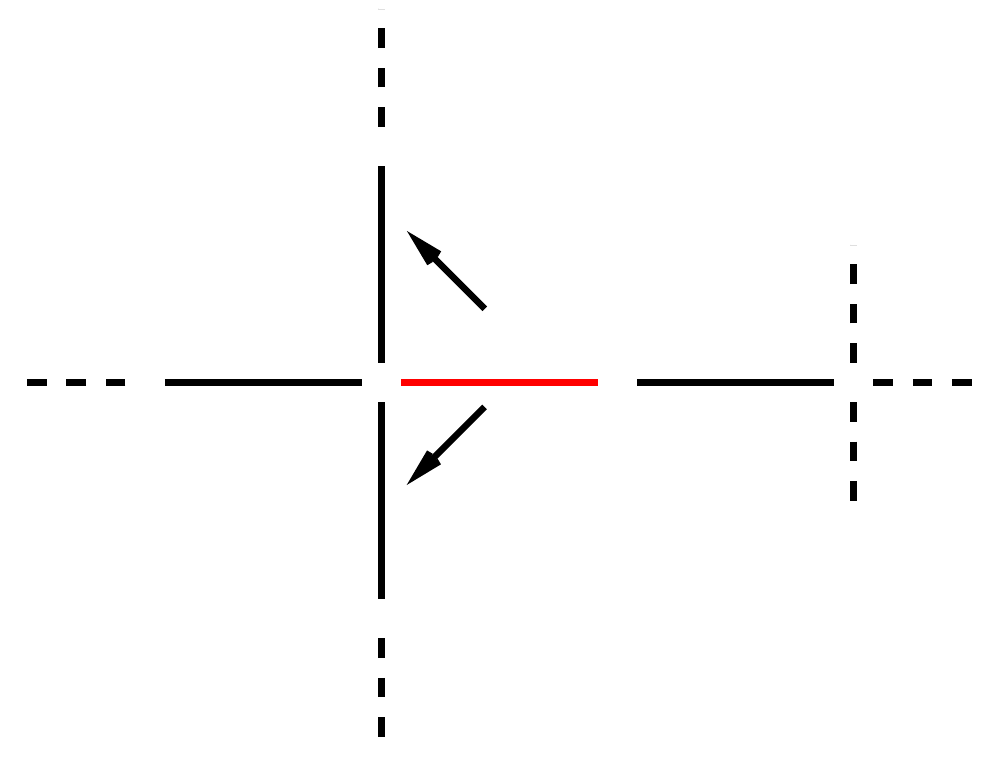
\caption{As an example, a qubit in a $\ket{0}$ eigenstate (red) is added to a toric code containing the plaquette (1,2,3,4,5,6) which gets split in two in the process. The toric code grows by one plaquette (left). The arrows are CNOT operations with the target bit at the tip of the arrow. The links are numbered as in the main text. Another example of a qubit in a $\ket{+}$ eigenstate (red) added as a new star (right).}
\label{fig:addplaquette}
\end{figure}
To add or remove a qubit or a link to or from a plaquette or star, respectively, the $\cnot$ gate is applied to all the qubits which are part of the old or new plaquette or star, respectively (cf. fig. \ref{fig:addplaquette}). This changes the stabilizer condition such that the qubit in question becomes or ceases to be part of the stabilizer. First, it is useful to understand the commutation of CNOT gates with the Pauli matrices involved in the definition of the stabilizer.

We first explain how to add a plaquette. Let $p_{\text{old}}$ be a plaquette consisting of the links $1, \dots, i$. The check operator for this plaquette is
\begin{equation}
 \checkp_{p_{\text{old}}} = \pauli{z}_1 \pauli{z}_2 \dots \pauli{z}_i \quad \text{with} \quad  \checkp_{p_{\text{old}}} \ket{\stabil} = \ket{\stabil} \;.
\end{equation}
The $\cnot$ operation acting on control qubit $\ket{c}$ and target qubit $\ket{t}$ can be written in terms of the spin matrices as
\begin{equation}
 \cnot_{c \otimes t} = \frac12 \left[ (\id + \pauli{z} ) \otimes \id  + (\id - \pauli{z}) \otimes \pauli{x} \right] \;,
\end{equation}
where the first operator acts on the control bit and the second on the target bit.
In this notation, it can be easily checked that $\cnot$ acts by conjugation in the sense that the following hold\
\begin{eqnarray}
 \cnot \left( \id \otimes \pauli{z} \right) \cnot &=& \pauli{z} \otimes \pauli{z} \label{eq:conjzchange} \\
 \cnot \left( \pauli{z} \otimes \id \right) \cnot &=& \pauli{z} \otimes \id \\ 
 \cnot \left( \pauli{x} \otimes \id \right) \cnot &=& \pauli{x} \otimes \pauli{x} \label{eq:conjxchange}\\
 \cnot \left( \id \otimes \pauli{x} \right) \cnot &=& \id \otimes \pauli{x} \;.
\end{eqnarray}
Applied to a stabilized state as depicted in fig. \ref{fig:addplaquette} this means in particular
\begin{eqnarray}
 \cnot_{17} \cnot_{27} \cnot_{37} \left( \id_{123} \otimes \pauli{z}_7 \right) \ket{\stabil \otimes 0} &=& \left( \pauli{z}_{123} \otimes \pauli{z}_7 \right) \cnot_{123,7} \ket{\stabil \otimes 0} \;, \\
 \cnot_{123,7} \left( \id \otimes \pauli{x} \right) \ket{+ \otimes \stabil} &=& \left( \pauli{x} \otimes \pauli{x} \right) \cnot_{123,7} \ket{+ \otimes \stabil} \;,
\end{eqnarray}
where $\pauli{x}\ket{+}=\ket{+}$ is the $+1$ eigenstate of the $\pauli{x}$ operator. This way of adding a new qubit by creating a new plaquette typically divides an old plaquette, which means that some of its links make up one new plaquette operator and the others the second. The $\cnot$ operators act on either set of links of the old plaquette which can be arbitrarily chosen ($1\dots 3$ or $4 \dots 6$ in the example). We have adopted $\cnot_{123,7}$ as a short hand notation for a product of $\cnot$ operators acting on qubit $1,2,3$ as control bits while 7 is the target for each factor.
Now assume that $\ket{\stabil}$ is a stabilizer state. Hence
\begin{eqnarray}
 \cnot \left( \id \otimes \pauli{z} \right) \ket{\stabil \otimes 0} &\overset{\mathrm{eigenstate}}{=}& \cnot \ket{\stabil \otimes 0} \\
 &\overset{\eqref{eq:conjzchange}}{=}& \left( \pauli{z} \otimes \pauli{z} \right) \cnot \ket{\stabil \otimes 0} \;,
\end{eqnarray}
where the first equality holds as $\pauli{z} \ket{0} = \ket{0}$ and the second one holds due to conjugation. Therefore, the state $\ket{\stabil_{\text{new}}}=\cnot \ket{\stabil \otimes 0}$ is a stabilizer state for a stabilizer on a plaquette, containing all the links in $\checkp_{\text{old}}$ and the new link, just added to the toric code, such that
\begin{equation}
 \checkp_{\text{new}} \ket{\stabil_{\text{new}}} = \checkp_{\text{p}_{\text{part}}} \otimes \pauli{z} \ket{\stabil_{\text{new}}} = \ket{\stabil_{\text{new}}} \;.
\end{equation}
The check operator on the other new plaquette is automatically satisfied because $\stabil$ is a stabilizer of the old toric code such that applying the old plaquette operator on one part or the other yields the same result, either $+1$ or $-1$.
Similarly, to add a star to the toric code, we follow the same reasoning for 
\begin{eqnarray}
 \cnot \pauli{x} \otimes \id \ket{+ \otimes \stabil} &\overset{\mathrm{eigenstate}}{=}& \cnot \ket{+ \otimes \stabil} \\
 &\overset{\eqref{eq:conjxchange}}{=}& \pauli{x} \otimes \pauli{x} \cnot \ket{+ \otimes \stabil} \;,
\end{eqnarray}
such that now $\ket{\stabil_{\text{new}}}= \cnot \ket{+ \otimes \stabil}$ is a stabilizer state for
\begin{equation}
 \checks_{\text{new}} \ket{\stabil_{\text{new}}} =  \pauli{x} \otimes \checks_{\text{old}} \ket{\stabil_{\text{new}}} = \ket{\stabil_{\text{new}}} \;.
\end{equation}
Again, the stabilizer condition on the other new star is automatically fulfilled as guaranteed by the stabilizer condition of the old toric code.
Here, to add a qubit to a plaquette of an existing toric code, the qubit needs to be prepared in the $\ket{0}$ eigenstate first. Conversely, to add a qubit to a star it needs to be prepared in the $\ket{+}$ eigenstate.

By reversing the argument, qubits are removed from a toric code by the action of $\cnot$ gates, where the removed qubits are projected onto the $\ket{0}$ and $\ket{+}$ eigenstates, respectively.
Consider a state $\ket{\stabil\subsc{old}}=\ket{\stabil\subsc{new}\otimes\stabil\subsc{remove}}$ which fulfills the stabilizer conditions such that $\pauli{x}\otimes\pauli{x}\ket{\stabil\subsc{old}} = \ket{\stabil\subsc{old}}$ and $\pauli{z}\otimes\pauli{z}\ket{\stabil\subsc{old}} = \ket{\stabil\subsc{old}}$. To remove a plaquette, we realize that
\begin{eqnarray}
 \cnot \pauli{z}\otimes\pauli{z} \ket{\stabil\subsc{new}\otimes\stabil\subsc{remove}} &=& \cnot \ket{\stabil\subsc{new}\otimes\stabil\subsc{remove}} \\
 &=& \id \otimes \pauli{z} \cnot \ket{\stabil\subsc{new}\otimes\stabil\subsc{remove}} \;.
\end{eqnarray}
Since the $\pauli{z}$ in the last line acts on $\ket{\stabil\subsc{remove}}$, this means that the application of a $\cnot$ gate to a stabilized plaquette projects the target bit into $\ket{\stabil\subsc{remove}}= \ket{0}$ while $\cnot\pauli{z}\otimes\id\ket{\stabil\subsc{new}\otimes\stabil\subsc{remove}} = \pauli{z}\otimes\id \ket{\stabil\subsc{new}\otimes\stabil\subsc{remove}}$ and $\ket{\stabil\subsc{new}}$ is left untouched and is thus a stabilized state for the toric code with $\ket{\stabil\subsc{remove}}$ removed. Analogously, for removing a star, we use
\begin{eqnarray}
 \cnot \pauli{x}\otimes\pauli{x} \ket{\stabil\subsc{remove}\otimes\stabil\subsc{new}} &=& \cnot \ket{\stabil\subsc{remove}\otimes\stabil\subsc{new}} \\
 &=& \pauli{x}\otimes\id \cnot\ket{\stabil\subsc{remove}\otimes\stabil\subsc{new}} \;,
\end{eqnarray}
which means that the $\cnot$ operation projects its control bit into $\ket{\stabil\subsc{remove}} = \ket{+}$ while $\cnot \id\otimes\pauli{x} \ket{\stabil\subsc{remove}\otimes\stabil\subsc{new}} = \id\otimes\pauli{x} \ket{\stabil\subsc{remove}\otimes\stabil\subsc{new}}$. Again, after the action of the $\cnot$ gate, the stabilizer condition is satisfied on the state $\ket{\stabil\subsc{new}}$. $\cnot$ operations can be used to add qubits to the toric code in the sense that the new qubits are needed to fulfill the stabilizer conditions after their operation. They remove qubits from a toric code in the sense that the qubits must not be included to fulfill the stabilizer conditions afterwards.

\subsection{Adding and removing excited states}
\label{subsec:addrem_excited}

The stabilizer conditions determine the ground state of the toric code model. Excited states violate the stabilizer conditions such that they are eigenstates of the check operators with eigenvalue $-1$ instead of $+1$. Still, excited states can be added to and removed from the toric code.

\begin{figure}
\centering
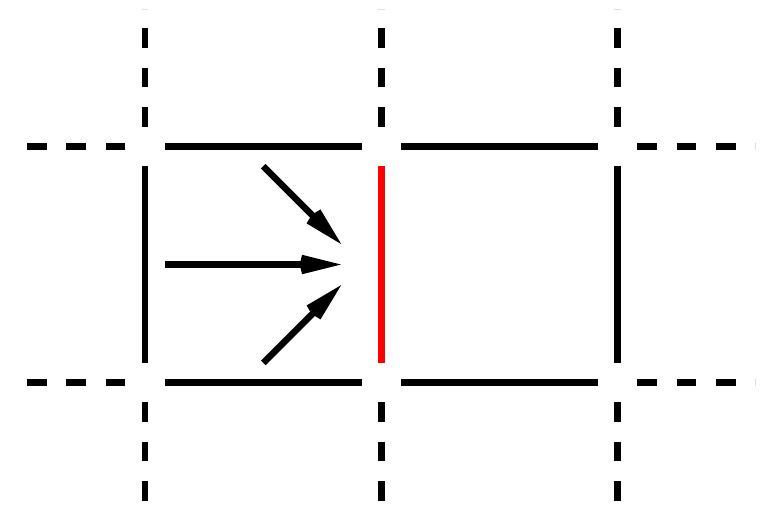
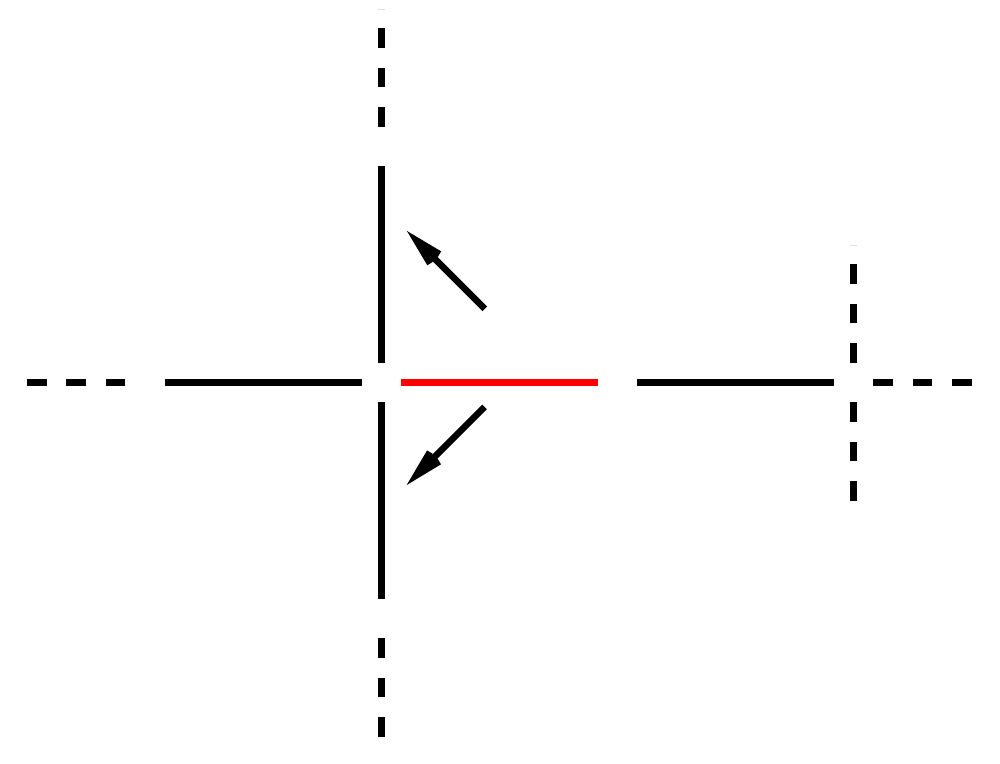
\caption{An example of a qubit in a $\ket{1}$ eigenstate (red) added to an existing toric code such that it grows by one plaquette (left). The arrows are CNOT operations with the target bit at the tip of the arrow. The links are numbered as in the main text. Another example of a qubit in a $\ket{-}$ eigenstate (red) added as a new plaquette (right).}
\label{fig:addflux}
\end{figure}
If the new qubit is prepared in a $\ket{1}$ state in the procedure for adding a plaquette, both plaquettes to which it belongs carry an excitation at the end. This is because applying the $\cnot$ operations then yields
\begin{eqnarray}
 \cnot \left( \id \otimes \pauli{z} \right) &=& - \cnot \ket{\stabil \otimes 1} \\
 &=& \left( \pauli{z} \otimes \pauli{z} \right) \cnot \ket{\stabil \otimes 1}
\end{eqnarray}
for either of the new plaquettes,
which just means that applying the new plaquette operator to an excited state indeed yields $\checkp\subsc{new} \ket{Z\subsc{new}} = - \ket{Z\subsc{new}}$.
Analogously, excited stars can be added to the toric code by preparing a new qubit in a $\ket{-}$ state
\begin{eqnarray}
 \cnot \pauli{x} \otimes \id \ket{- \otimes \stabil} &=& - \cnot \ket{- \otimes \stabil} \\
 &=& \pauli{x}\otimes\pauli{x} \cnot \ket{- \otimes \stabil} \;.
\end{eqnarray}

Correspondingly, removing an excited plaquette or star from a toric code projects the qubit to be removed into a $\ket{1}$ or $\ket{-}$ state, respectively. For an excited plaquette, the check operator yields a $(-1)$ eigenstate
\begin{eqnarray}
 \cnot \pauli{z}\otimes\pauli{z} \ket{\stabil\subsc{new}\otimes\stabil\subsc{remove}} &=& -\cnot \ket{\stabil\subsc{new}\otimes\stabil\subsc{remove}} \\
 &=& \id \otimes \pauli{z} \cnot \ket{\stabil\subsc{new}\otimes\stabil\subsc{remove}} \;,
\end{eqnarray}
such that $\ket{\stabil\subsc{remove}} = \ket{1}$. For removing an excited star
\begin{eqnarray}
 \cnot \pauli{x}\otimes\pauli{x} \ket{\stabil\subsc{remove}\otimes\stabil\subsc{new}} &=& - \cnot \ket{\stabil\subsc{remove}\otimes\stabil\subsc{new}} \\
 &=& \pauli{x}\otimes\id \cnot\ket{\stabil\subsc{remove}\otimes\stabil\subsc{new}}
\end{eqnarray}
and $\ket{\stabil\subsc{remove}}$ gets projected to $\ket{-}$. 

\begin{figure}
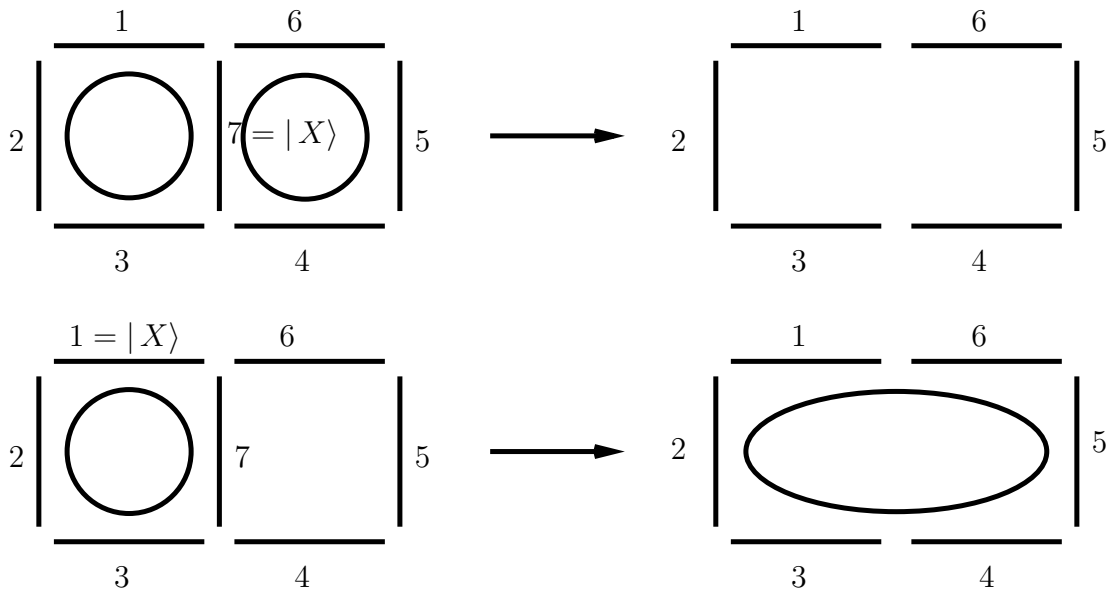
\label{fig:removeplaquettewithflux}
\caption{When removing a plaquette/star with an excitation, the excitations annihilate if the removed link was carrying the excitation. Otherwise, it survives and spreads onto the new plaquette/star.}
\end{figure}

Notice that there are two possible situations when removing a qubit from a plaquette or a star carrying an excitation: Either, the qubit removed is the one on which the excitation pair hinges (i.e. the one to which the $\pauli{x}$ or $\pauli{z}$ had been applied), or the removed qubit is non-essential for the excitation. In both cases, the stabilizer condition on the old plaquette or star, respectively, is violated and thus the removed qubit is projected into the $\ket{1}$ or $\ket{-}$ state, irrespective of its r\^{o}le played for the excitation. However, in the first case, the excitation vanishes with the qubit while in the second case, the excitation lives on the new plaquette after removal. We can see this by examining the check operators after removal. Let us consider the two adjacent plaquettes $1234$ and $4567$ as depicted in fig. \ref{fig:removeplaquettewithflux} where the excitation hinges on qubit $4$. 
Removing this qubit yields
\begin{eqnarray}
 \cnot_{14}\cnot_{24}\cnot_{34} \pauli{z}_1 \otimes \pauli{z}_2 \otimes \pauli{z}_3 \otimes \id_4 \ket{\stabil} &=& \pauli{z}_1 \otimes \pauli{z}_2 \otimes \pauli{z}_3 \otimes \id_4  \cnot_{14}\cnot_{24}\cnot_{34} \ket{\stabil} \nonumber \\
 \cnot_{54}\cnot_{64}\cnot_{74} \pauli{z}_5 \otimes \pauli{z}_6 \otimes \pauli{z}_7 \otimes \id_4 \ket{\stabil} &=& \pauli{z}_5 \otimes \pauli{z}_6 \otimes \pauli{z}_7 \otimes \id_4  \cnot_{54}\cnot_{64}\cnot_{74} \ket{\stabil} \nonumber \;.
\end{eqnarray}
From the check operators, we know
\begin{eqnarray}
 \checkp_{1234} \ket{\stabil} &=  \pauli{z}_1 \otimes \pauli{z}_2 \otimes \pauli{z}_3 \otimes  \pauli{z}_4 \ket{\stabil} & = - \ket{\stabil} \\
 \checkp_{5674} \ket{\stabil} &=  \pauli{z}_5 \otimes \pauli{z}_6 \otimes \pauli{z}_7 \otimes  \pauli{z}_4 \ket{\stabil} & = - \ket{\stabil} \;.
\end{eqnarray}
Applying $\pauli{4}$ to both sides, we seen
\begin{equation}
 \pauli{z}_1 \otimes \pauli{z}_2 \otimes \pauli{z}_3 \otimes  \id_4 \ket{\stabil} = - \pauli{z}_4 \ket{\stabil} = \pauli{z}_5 \otimes \pauli{z}_6 \otimes \pauli{z}_7 \otimes  \id_4 \ket{\stabil} \;.
\end{equation}
Consequently, the check operator on the new plaquette
\begin{equation}
\checkp_{123567} \cnot \ket{\stabil} = \pauli{z}_1 \otimes \pauli{z}_2 \otimes \pauli{z}_3 \otimes  \id_4 \otimes \pauli{z}_5 \otimes \pauli{z}_6 \otimes \pauli{z}_7 \cnot \ket{\stabil} = (-1)^2 \cnot \ket{\stabil} 
\end{equation}
and thus the new toric code state satisfies the new stabilizer condition and the excitation has vanished.
If, on the other hand, the qubit which carries the excitation, qubit $1$ say, is not removed the old check operators yield
\begin{eqnarray}
 \pauli{z}_1 \otimes \pauli{z}_2 \otimes \pauli{z}_3 \otimes  \id_4 \ket{\stabil} &=& - \pauli{z}_4 \ket{\stabil} \\
 \pauli{z}_5 \otimes \pauli{z}_6 \otimes \pauli{z}_7 \otimes  \id_4 \ket{\stabil} &=& \pauli{z}_4 \ket{\stabil} \;,
\end{eqnarray}
from which we see that the new check operator
\begin{equation}
 \checkp_{123567} \cnot \ket{\stabil} = - \cnot \ket{\stabil}
\end{equation}
and one flux now lives on the new, bigger plaquette. 

Another observation to be made is that the removal of a plaquette preserves the number of the stars and thus potential charge excitations. The number of rays of the star changes, though. Two neighboring charges only annihilate if they are made to coincide due to star removal. Analogously, removal of a star preserves the number of plaquettes and thus also their flux excitations, although the circumference of the plaquette next to the star changes.

\section{Extension of MERA to represent excited states}
\label{sec:excited}

Using the method in subsection \ref{subsec:addremove}, disentanglers and coarse grainers can be constructed for the toric code analytically and explicitly as has been presented in \cite{Aguado2008}. They only contain CNOT operations, where certain qubits are projected into the $\ket{0}$ and $\ket{+}$ states. These states are effectively removed from the toric code in the sense that the stabilizer conditions are fulfilled on the plaquettes and stars, respectively, without those qubits. For details of this construction we refer the reader to \cite{Aguado2008}.

MERA gives a representation of the quantum state of the underlying system if the whole network is considered: the quantum state is characterized by recording the spins which are being traced out by the coarse graining operations. As such, the network is also capable of keeping track of excitations. Following a similar procedure as outlined in section \ref{subsec:addrem_excited}, it is obvious that for excited states the coarse grainer outputs $\ket{1}$ and $\ket{-}$ states for the plaquettes and stars carrying excitations in this construction.

\subsection{Stretching of excitations}

The flux and charge excitation of the toric code need not be localized. If created by switching one link the fluxes reside on the plaquettes or stars adjacent to the flipped qubit. But since the excitations consist of two particles which are their own anti-particles, applying the same flip once again to a link adjacent to one of the particles, it gets shifted to the next plaquette or star, respectively. The stabilizer condition is restored on the intermediate plaquette. By applying such operators on adjacent links, a string operator is created, which separates the excitations to an arbitrary distance until the size of the toric code lattice. Note that if a string operator spans over a whole grand cycle of the torus, the ground state and thus the whole manifold of states is changed into a different sector. We will at first not consider such a topological change but only different scales of excitations within one manifold.

In the construction of MERA for the toric code, the relevant degrees of freedom are the plaquettes and stars, not individual qubits. Therefore, the MERA is sensitive to the excited plaquettes itself and does not see if a string operator connecting them is acting on qubits in between.

The effect of the real space renormalization on the excitation depends on where the excitations live. When after applying the simultaneous qubit removal operations two excitations coincide, they cancel each other as explained above. The MERA as suggested in \cite{Aguado2008} takes 4 meta-sites of 4 qubits, which are reduced to 1 qubit each by the disentangling and coarse graining procedures together. Correspondingly, 8 plaquette and star degrees of freedom respectively are reduced to two plaquettes and 2 stars. Excitations can be integrated out in the disentangling step, in the coarse graining step or not at all. The latter happens if the excitations are distributed such that the removals do not make them coincide. Excitations are not necessarily integrated out if they live within one meta-site.

\begin{figure}
 a) \includegraphics[width=0.2\textwidth]{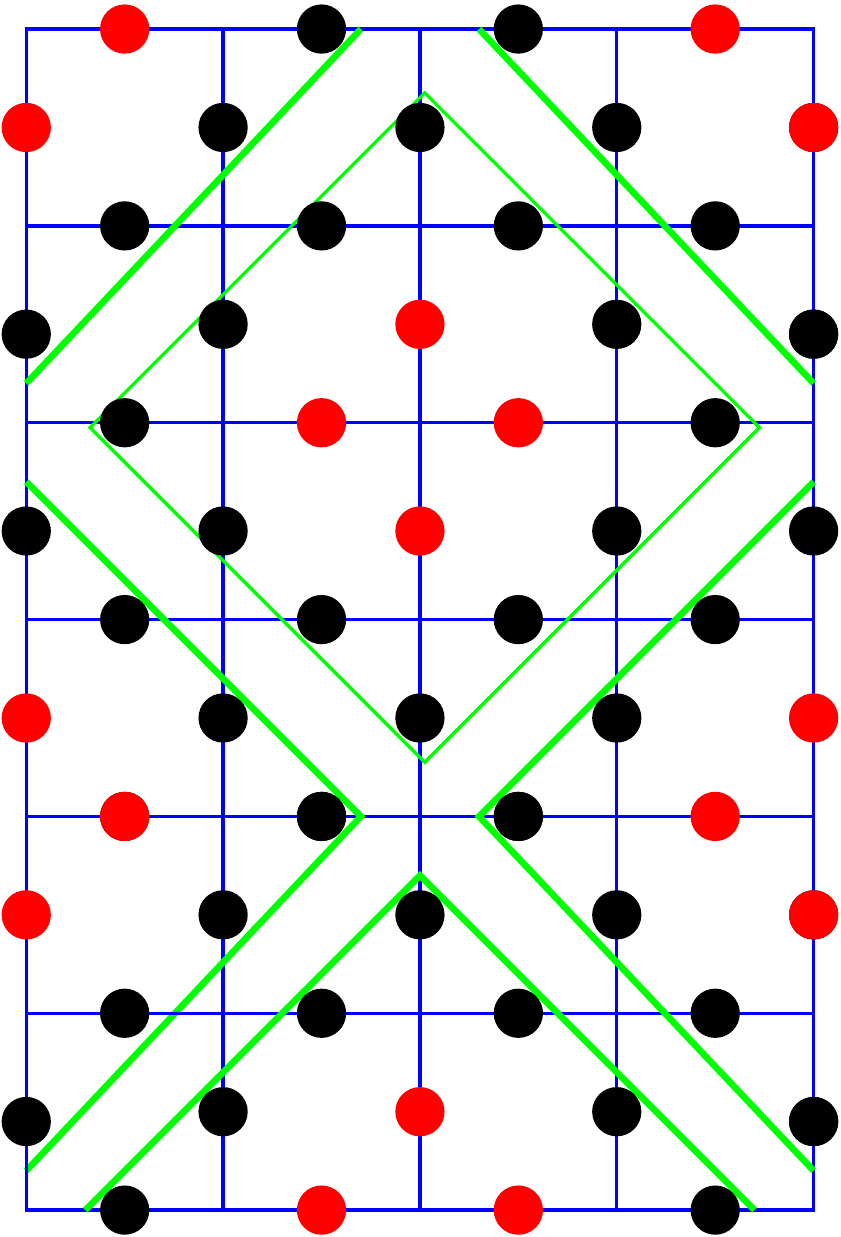}\vspace{0.1\textwidth}
 b) \includegraphics[width=0.2\textwidth]{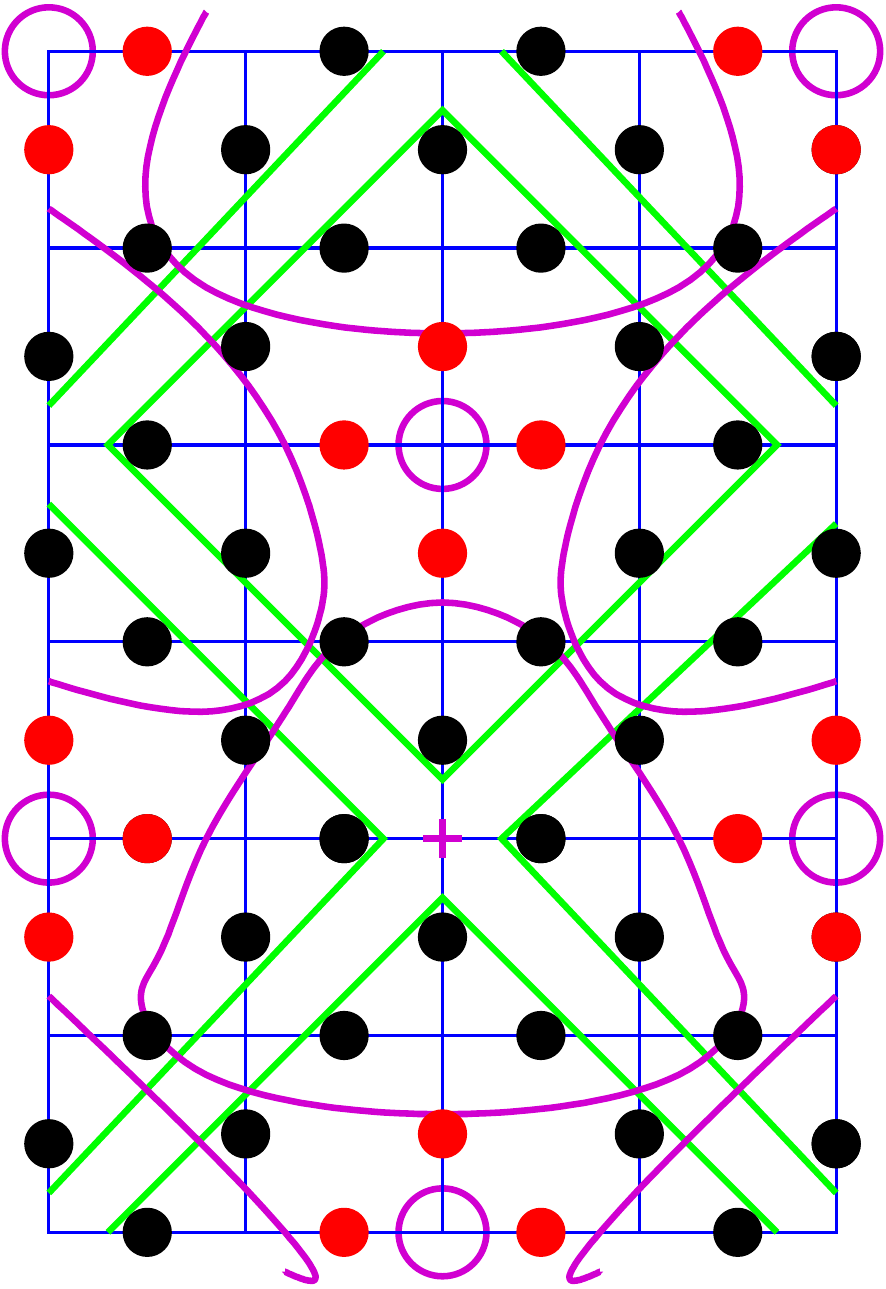}\vspace{0.1\textwidth}
 c) \includegraphics[width=0.2\textwidth]{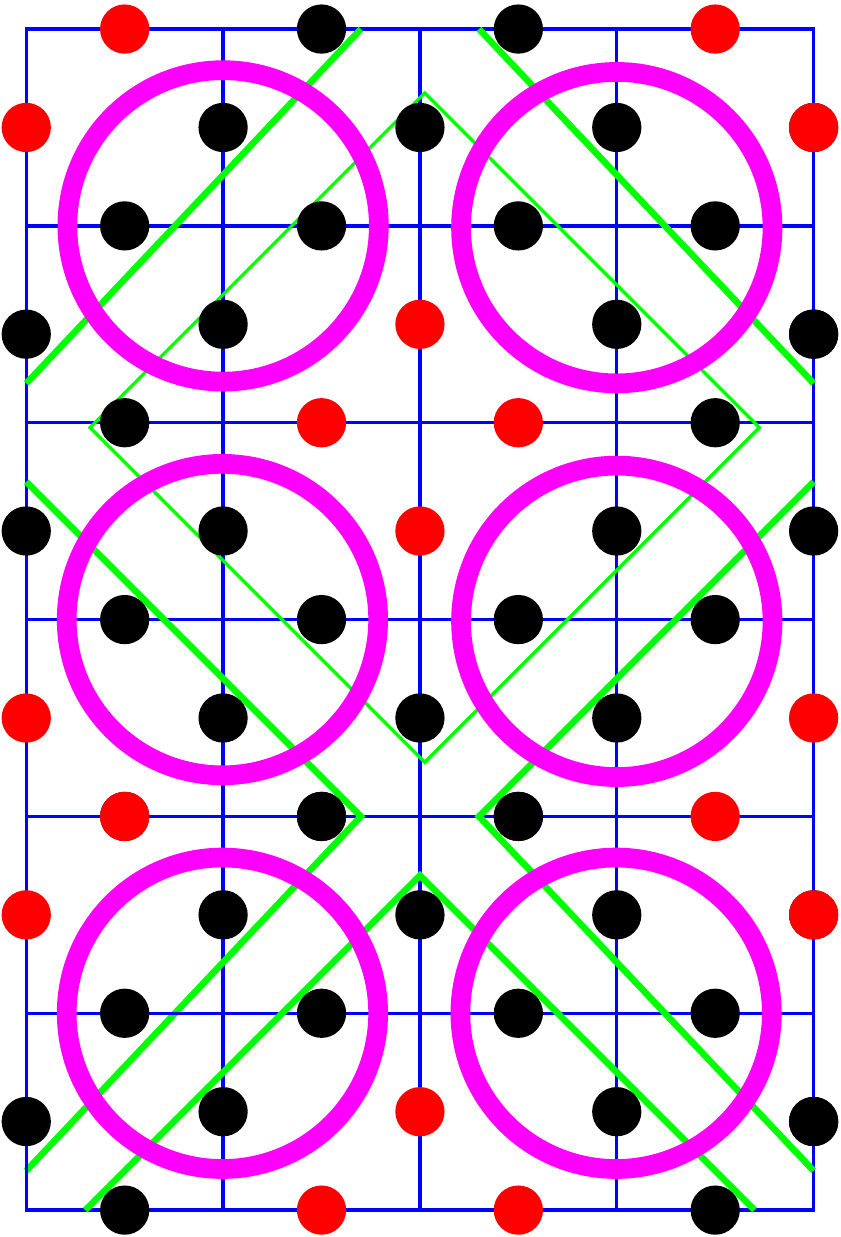}
 \caption{a) The MERA of the toric code according to \cite{Aguado2008} is defined for 4 meta-sites of 4 qubits each, which are delimited with the green line above. The qubits which remain after both disentangling and coarse graining steps are colored in red. b) The RG reduces the star degrees of freedom by a factor 4 just like the qubits. The stars delimited by the purple curve all combine into one star after a full RG step. This means that all charge excitation pairs in this area annihilate after one RG-step. The star in the middle of the meta-site, which is encircled in purple, remains unchanged and a charge on this site survives. c) Accordingly, also four plaquettes combine into one after an RG step. Those are denoted with a purple circle. Again, flux pairs living on those four plaquettes annihilate.}
 \label{fig:meraprocedure}
\end{figure}

After the isometry, each four adjacent plaquettes forming a square are made to coincide (see fig. \ref{fig:meraprocedure} a), so fluxes living on those plaquette cancel each other. One meta-site stretches over four plaquettes. As for the stars, the one in the middle of each meta-site remains untouched completely by the renormalization procedure, while seven stars bridging between the meta-sites are combined to one (see fig. \ref{fig:meraprocedure} b).
The notion of locality preservation of the renormalization procedure entails that excitations should be integrated out on an energy scale inversely proportional to their spatial separation, i.e. excitations close to each other close to the boundary and far excitations apart from each other deep in the bulk. For the plaquettes, excitations localized on the four adjacent ones are combined into one at each renormalization step. It is easy to see that those fluxes which live on adjacent plaquettes but end up on different plaquettes after renormalization are combined into one after the next renormalization step, which is an effect of discretization. For the charges, we see that the star in the middle of each meta-site remains untouched, while the ones at the corners of a meta-site are combined with their neighbours in a cross-like manner. At the next renormalization step, 16 remaining qubits are combined to a new meta-site around a star, which was in between the meta-sites of the previous renormalization step, when it was left untouched. Thus, from step to step, the r\^{o}le of the relative location of the stars is mixed and they get combined with their neighbours at the next renormalization step. The notion of locality is thus preserved by the renormalization procedure in the sense that excitations which are close at each coarse graining step are integrated out. 

We make a final remark concerning the degeneracy and the scale of the excitations: as we go deeper into the MERA the number of degrees of freedom changes and so does the degeneracy of excitations according to \eqref{eq:deg_excitation}. Obviously, there are less possibilities to distribute the two particles of excitations further apart such that they survive longer in the renormalization procedure, which corresponds to a smaller degeneracy of the toric code at lower energy scales. Also note that the coupling constants $\lambda_{p,s}$ in \eqref{eq:Ham_tc} are given in terms of the inverse lattice constant $1/a$, which decreases in each RG-step by a factor 2. This also applies for the energy of each excitation, which is $4 \lambda_{p,s}$, respectively.

\section{Topological entanglement entropy in MERA}
\label{sec:tee}

The principal feature of the toric code is its topological order. It has been shown \cite{Kitaev2006} that this order can be detected in the ground state from a constant $\gamma$ in the entanglement entropy
\begin{equation}
 S(\rho) = \alpha L - \gamma + \mathcal{O}(\frac{1}{L}) \;,
\end{equation}
where the first term comes from the entanglement across the boundary of the subsystem and is therefore proportional to it, while terms which are omitted vanish in the limit $L \to \infty$. The so-called topological long-range entanglement entropy $\gamma$ characterizes topological entanglement of the ground state -- and in the case of the toric code also of arbitrary excited states. The total quantum dimension of the system $\mathcal{D}$ is related to $\gamma = \log \mathcal{D}$, where $\mathcal{D} = \sqrt{\sum_a d_a^2}$ and $d_a$ is the quantum dimension of a particle with charge $a$. For abelian anyons $d_{\mathrm{abelian}}=1$ and hence the toric code has $\gamma = \log 2$ \cite{Kitaev2006,Hamma2005}.

\begin{figure}
 \centering
 \includegraphics[width=0.3\textwidth]{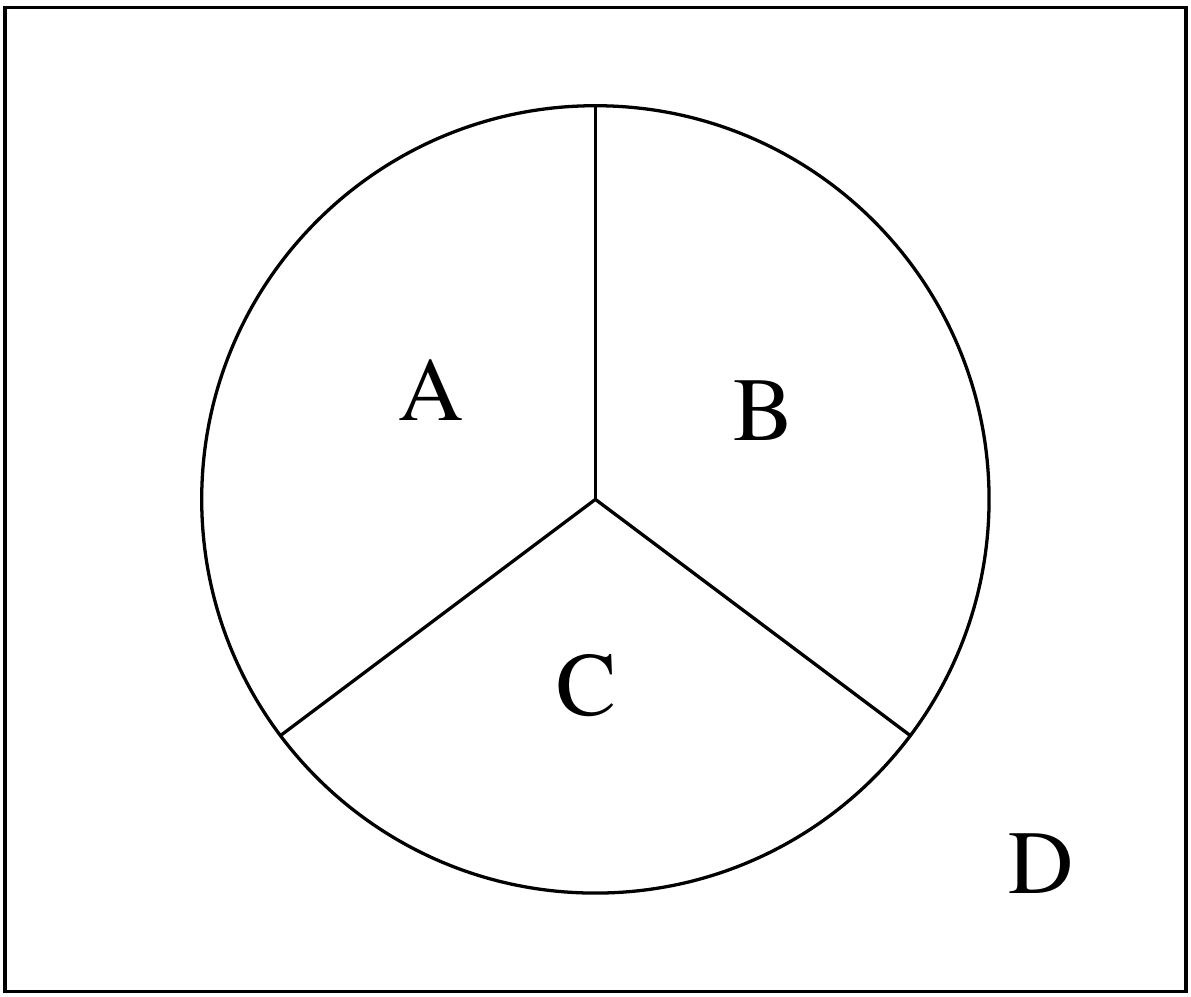}
 \hspace{0.05\textwidth}
 \includegraphics[width=0.4\textwidth]{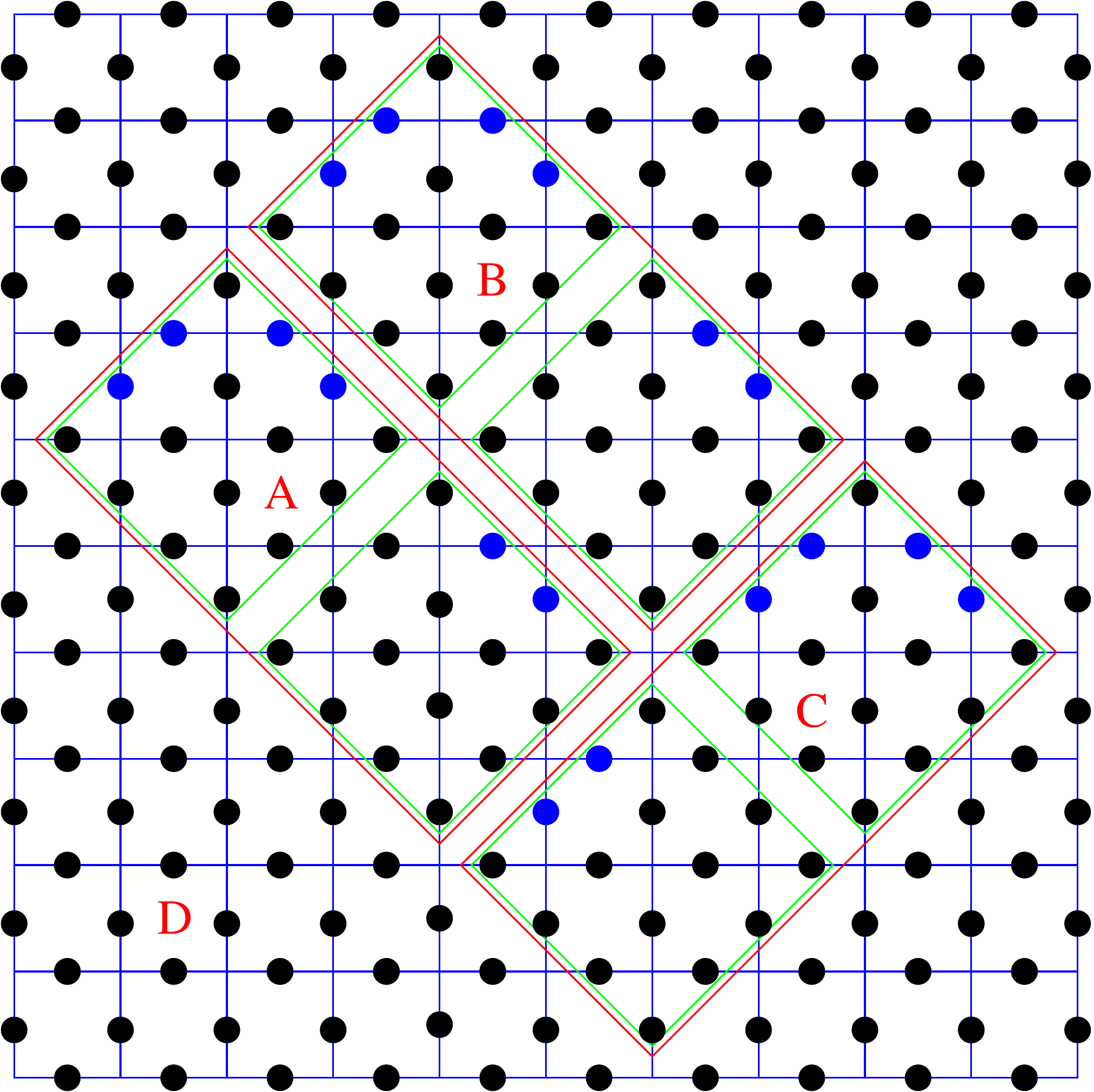}
 \caption{The topological contribution to the entanglement entropy can be calculated from a suitable combination of subsystems (left) and an example in the toric code (right). The regions delimited in red represent the smallest subsystems $A,B,C$. The sites delimited in green form a meta-site of the MERA procedure. (color online)}
 \label{fig:tee_local}
\end{figure}

The universal constant $\gamma$ can also be determined locally by combining the entanglement entropies of adjacent subsystems in a suitable way such that the contributions from their boundaries cancel. For a two-dimensional system like the toric code, the topological entanglement entropy \cite{Kitaev2006}
\begin{equation}
S_{\mathrm{topo}} = -\gamma = S_A + S_B + S_C - S_{AB} - S_{BC} - S_{AC} + S_{ABC} \;,
\label{eq:Stopo}
\end{equation}
where the subsystems are defined as in fig. \ref{fig:tee_local}. Equation \eqref{eq:Stopo} is valid as long as the size of the individual subsystems is larger than the correlation length in the system. For the toric code, the correlation length vanishes conveniently, so that this condition is always met.
A concrete possibility to calculate the topological entanglement entropy for the toric code is depicted in Fig. \ref{fig:tee_local}.

In the following, we review how the entanglement entropy is calculated in a holographic theory as a minimal surface and then we are going to present how to obtain the topological entanglement entropy from the MERA in the toric code, which generalizes to arbitrary systems.

\subsection{Entanglement entropy}

The entanglement entropy of a subsystem $A$ of the total system $A \cup B$ is the von Neumann entropy $S_A = - \tr \rho_A \log \rho_A$ of the reduced density matrix $\rho_A = \tr_B \rho$ with the degrees of freedom living in subsystem $B$ traced out. In a MERA representation of a quantum state of a system, the entanglement entropy is measured by the ``area of a surface'' cutting the minimal number of bonds such that the subsystem under consideration and its sub-network are completely separated from the rest of the system (cf. fig. \ref{fig:minimalcurve}). This property is reminiscent of the Ryu-Takayanagi formula \cite{Ryu2006}, which relates the entanglement entropy of a subsystem to the area of a minimal surface spanned by the boundary of the subsystem into the bulk. The surface area of the causal cone in MERA is measured by the number of links that pierce it. The total entanglement entropy is given by the sum of the entropy contributed by each layer in the MERA network, i.e.
\begin{equation}
 S_A = \sum_{i \in \{ \text{cut bonds} \} } d_i \;,
\end{equation}
where $d_i$ is the bond dimension of each leg to be cut, which is the dimension of the Hilbert space represented by a bond, which is removed in the RG step.

For stabilizer states and for excited states in which the plaquette check operators remain unchanged, the entanglement entropy is known to scale with the number of plaquette operators involved in the boundary between the subsystems \cite{Hamma2005,Castelnovo2007}. 
We can therefore compare our holographic result to the result from group theoretic considerations.

Following \cite{Hamma2005}, the entanglement entropies of the regions delimited in fig. \ref{fig:tee_local} are calculated from counting the number of sites $n_i$ outside the subsystem whose stars act on $i$ qubits inside the subsystem. The entanglement entropy is proportional to the length of the boundary 
\begin{equation}
 S_A = L_{\partial A} - n_2 -2 n_3 -1 \;,
\end{equation}
where $L_{\partial A} = n_1 + 2 n_2 + 3 n_3$, which leads to
\begin{equation}
 S_A = n_1 + n_2 + n_3 -1 \;.
\end{equation}
We have chosen the smallest subsystems (red in fig. \ref{fig:tee_local}) such that they comprise two sets of four meta-sites (green in fig. \ref{fig:tee_local}) following the procedure outlined in \cite{Aguado2008}.
In our case, $n_3=0$ for all the subsystems since their boundary is convex.
For the small patches $A,B,C$, we have 
\begin{equation}
  n_1^{A,B,C} = 4 \;, \quad n_2^{A,B,C} = 8 \;,
\end{equation}
and thus find for their entanglement entropy $S_A = S_B = S_C = 11$.
For the subsystems combining two of the small patches, the entanglement entropies for the different combinations are slightly different, because while $AB$ is convex, $BC$ and $AC$ are not. Hence we find
\begin{align}
 n_1^{AB} & = 4 \;, & n_2^{AB} & = 12 \;, & n_3^{AB} & = 0 \;,
 n_1^{AC} &= n_1^{BC} = 5 \;,  & n_2^{AC} & = n_2^{BC} = 14 \;, & n_3^{AC} &= n_3^{BC} = 1 \;,
\end{align}
and the entanglement entropies of the regions are $S_{AB}=15$ while $S_{AC} = S_{BC} = 19$. Finally, for the collection of all the sub-regions, we obtain
\begin{equation}
 n_1^{ABC} = 4 \;, \quad n_2^{ABC} = 16 \;, \quad n_3^{ABC} = 0 \;,
\end{equation}
so that the entanglement entropy is $S_{ABC} = 19$. We can now calculate the topological entanglement entropy according to \eqref{eq:Stopo} 
\begin{equation}
 \gamma = - S_{\mathrm{topo}} = 1 \;,
\end{equation}
which agrees with the value obtained from the quantum dimension of the system (We consider the logarithm with respect to basis 2, which is natural for qubits.).

\subsection{Geometric interpretation of the topological entanglement entropy}

Since we have established how to calculate the entanglement entropy of a finite subsystem, we can also evaluate the topological entanglement entropy \eqref{eq:Stopo} successively. Some features of this calculation are clear right away. 

\begin{figure}
 \centering
 \includegraphics[width=0.5\textwidth]{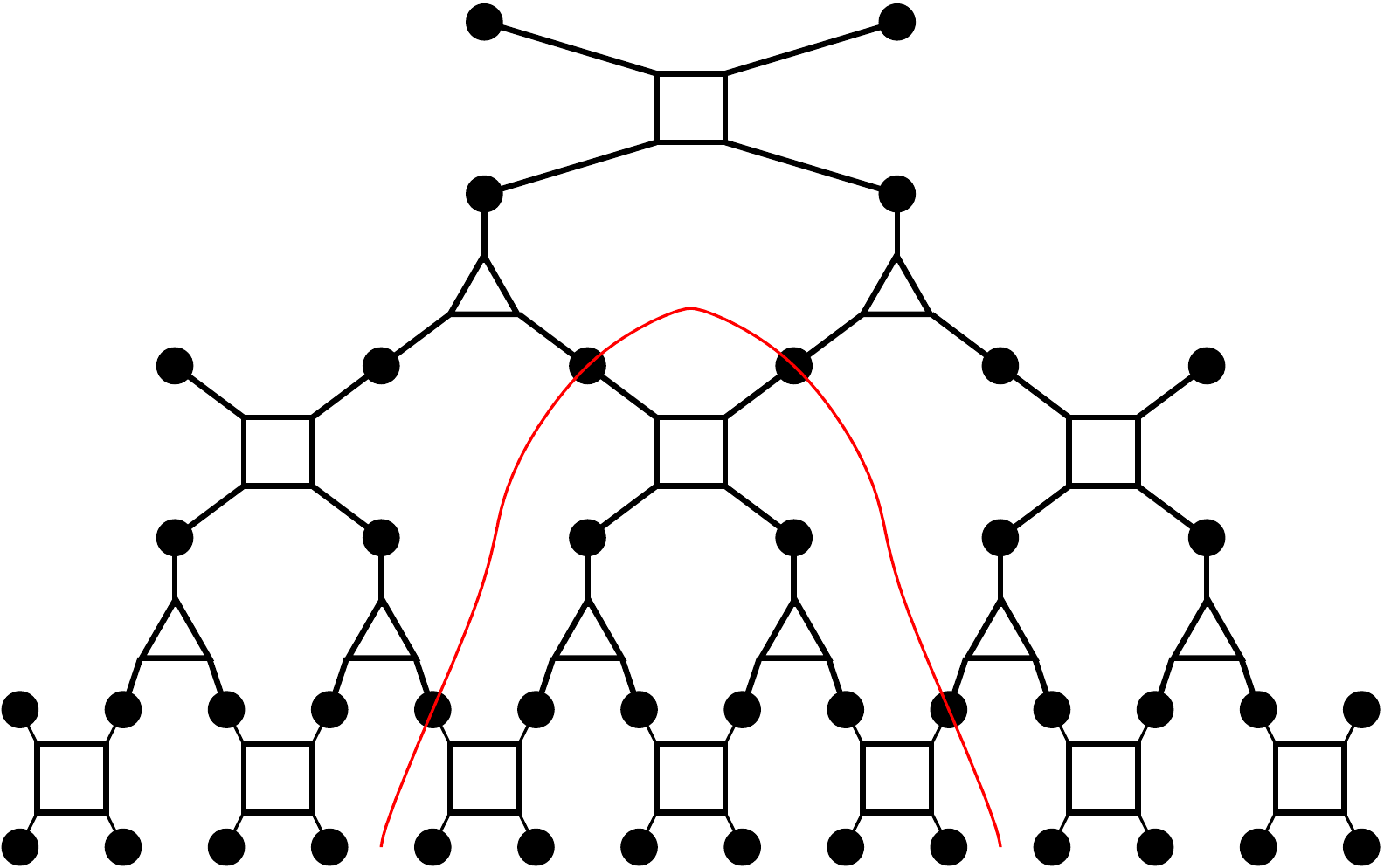}
 \caption{The entanglement entropy of a subsystem is bounded by the minimal number of links a curve needs to cut in the network in order to separate it from the rest of the system.}
 \label{fig:minimalcurve}
\end{figure}

In the following, we will re-obtain the value from a holographic calculation in MERA.
The calculation of the entanglement entropy in MERA resembles the procedure of calculating the entanglement entropy in the AdS/CFT correspondence. In the latter, the entanglement entropy is calculated following the Ryu-Takayanagi formula \cite{Ryu2006} as the area of a minimal surface stretching from the boundary of the region in question into the bulk. To obtain this bound, the induced metric on the surface is extremized. Correspondingly, in MERA the entanglement entropy turns out to be bound by the minimal number of links which are cut by a curve stretching into the network from the boundary of the region in the spin chain (see fig. \ref{fig:minimalcurve}).
For our case of a 2-dimensional system, the bonds are cut along a 2-dimensional surface extending into the network and separating a patch from the rest of the system. The entanglement entropy is again the sum of the dimensions of all the bonds cut in the procedure.

The dimension of each bond is the dimension of the Hilbert space of each qubit, hence $d=2$. 
Of course, the number of bonds to be cut depends on the size of the patches and on which bonds contribute to the entanglement entropy for the given renormalization procedure. However, as it will turn out, the bonds closer to the boundary are cut for all the patches and thus drop out of the final result. We begin our consideration again for the small patches $A,B,C$ in fig. \ref{fig:tee_local}. Again, the entanglement entropy for each of these is the same. 

To obtain the correct  numerical value of the entanglement entropy we have to take into account how much entanglement a given bond removes from the system. In the case of the toric code, around each of the 4 by 4 sites, on which the disentanglers are defined, only the four qubits which are projected onto the $\pauli{z}$ eigenstates contribute. The entanglement entropy at the lowest level $\tau$ is then obtained by adding the entanglement removed with the entanglement entropy of the remaining system after coarse graining (which in itself does not remove any more entanglement from the system)
\begin{equation}
 S(\tau) = \sum_{\ket{i}\in\{\ket{0},\ket{1}\}} \log_2 \dim_{i} + S(\tau +1) \;,
\end{equation}
where $\dim_i$ is the dimension of the Hilbert space at the cutoff, which is $\dim_{\mathrm{qubit}}=2$ and the sum just counts the number of qubits projected out and cut through. If the number of qubits in the subsystem at level $\tau+1$ is larger than 4 by 4, we can perform another RG-step, otherwise we add the entanglement entropy of the remaining qubits. 

For the smallest patches $A,B,C$, this means that we collect six qubits along the edge of a 2 by 2 meta-site (blue qubits in fig. \ref{fig:tee_local}). After the coarse graining step, both 2 by 2 meta-sites are reduced to 2 by 2 qubits each so that the remaining entropy is that of a 4 by 2 patch, which is $S_{4 \times 2}= 5$ following the formula of \cite{Hamma2005}. This is not enough to perform another complete RG-step, for which we need a 4 by 4 patch. Hence, the entanglement entropies of the small patches are $S^A=S^B=S^C=6 + 5 = 11$. For the larger patch $AB$, we can perform two RG-steps. The first one contributes an entanglement of eight qubits around the boundary and the coarse graining step combines the 2 by 2 meta-sites into one meta-site, around which we collect again four qubits. After the second coarse graining procedure, we are left with a 2 by 2 subsystem, which has an entanglement entropy of $S_{2 \times 2}=3$ and the total entropy of this patch is $S^{AB}= 8 + 4 + 3 = 15$. For the two L-shaped patches $AC$ and $BC$, the disentanglers at the first step collect a contribution from 10 qubits. The coarse graining step combines the meta sites into an L-shaped with 2 by 2 qubits per meta-site. Around the disentangler on the second level, we collect 5 qubits and the second coarse grainers leave us with an L-shaped form of 4 by 2 qubits. This shape has an entanglement entropy of 4, which collects to $S^{AC}=S^{BC}=10 + 5 +4 =19$. Finally, for the biggest patch $ABC$, around the boundary we collect 10 qubits in the first step. Around the 4 by 6 patch in the second RG-step, we have 5 qubits contributing to the entanglement entropy, leaving us with a 2 by 3 patch upon coarse graining, which contributes an entanglement entropy of 4. In total, we collect $S^{ABC}=10+5+4=19$. All these geometric values of the entanglement entropy agree with the ones obtained earlier. 

\begin{figure}
\resizebox{0.9\textwidth}{!}{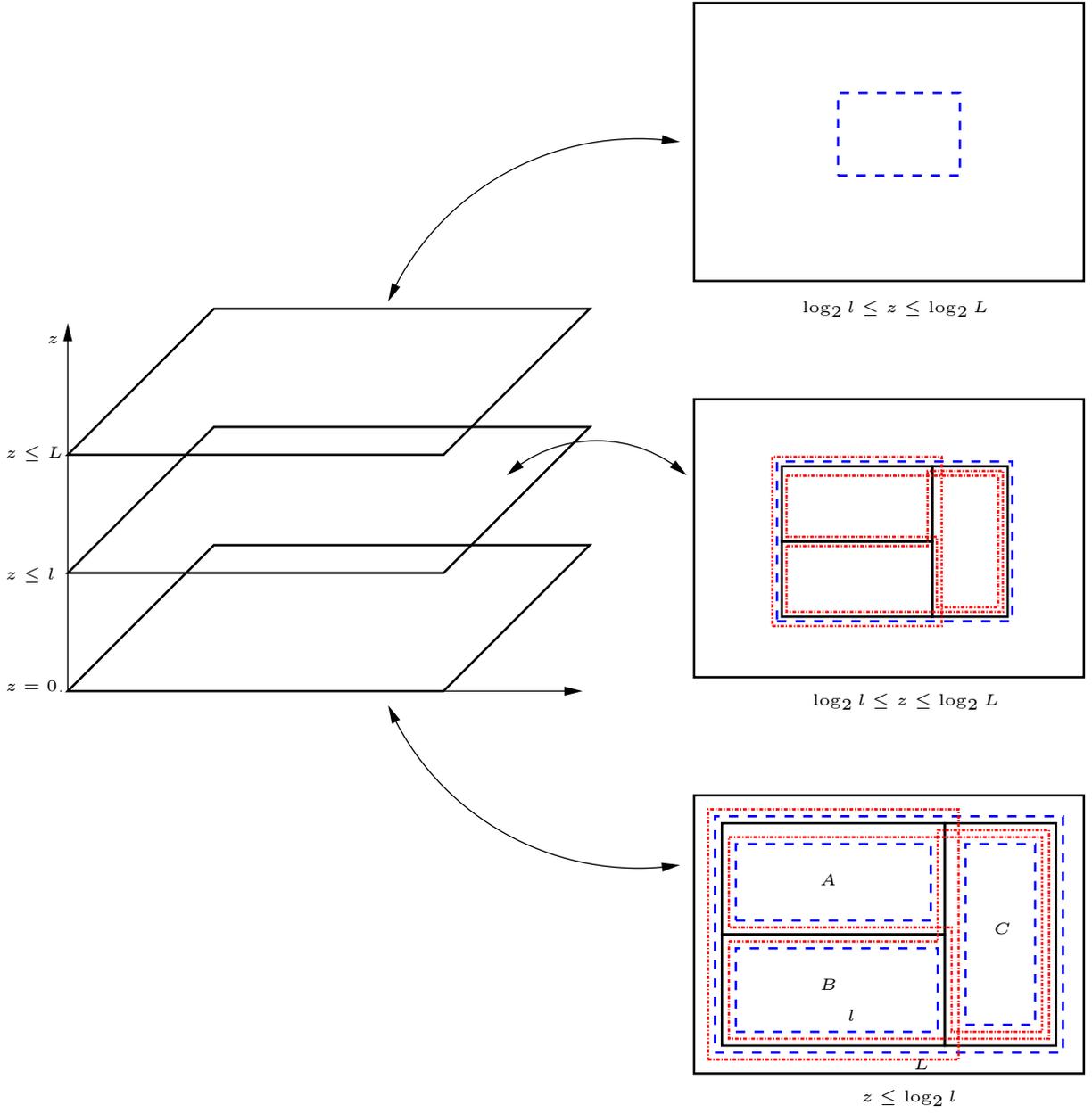}
\caption{The holographic contribution to the topological entanglement entropy on different scales. Solid lines (blue) denote bonds which are added, dotted lines (red) denote bonds which are subtracted according to \eqref{eq:Stopo}. In the UV at $z=0$ (leftmost), the contribution from all the boundaries cancels agains each other as they should. At the scale $z = \log_2 l$ (middle) the smallest patches with size $l$ (in terms of lattice points) have been coarsened away and do not contribute any more to the entanglement entropy. For simplicity, we only consider the scale of the patches in one dimension and combine the disentanglers and coarse grainers in both dimensions into one layer. At the scale $z= \log_2 L$ of the combined patch $ABC$ with size $L>l$, all (negative) contributions from the patches $AB$, $AC$ and $BC$ have been coarsened and the only contribution remaining comes from $ABC$. (Color online)}
\label{fig:holoEE_levels}
\end{figure}

However, the interpretation is now different: While the bonds which contribute to the entanglement entropy in the first steps of the RG-procedure, i.e. closer to the UV cancel in the construction of the topological entanglement entropy against contributions from other patches, the RG of the smaller patches stops earlier than the one of the large patches. Hence, deeper in the IR, there are bonds which only belong to the RG of a larger patch. Their contribution survives and the topological entanglement entropy comes from a geometrical ``defect'' located the deeper in the geometry, i.e. further in the IR, the larger the patches are which are used to measure the topological entanglement entropy (cf. fig. \ref{fig:holoEE_levels}). The topological quantity is therefore not localized at a specific depth in the geometry as is actually to be expected. 

Albeit the toric code is a model with rather particular properties, we stress that this geometric picture of measuring the topological entanglement entropy as a surface in the bulk is a generic one. The construction of Kitaev and Preskill \cite{Kitaev2006} is such that the short-range entanglement across the boundaries cancel, which are geometrically located in the UV close to the boundary ($z \approx 0$). The minimal surface, bounded by a patch in the lattice model stretches the deeper into the bulk, the bigger the individual patches are. The smaller patches with a size $l$ in terms of the lattice spacing $a$ are coarse grained into a single site at about an RG step at $z=\log_2 \frac{l}{a}$. Beyond that depth in the bulk, not all contributions from the different patches cancel against each other and the surface extending further remains and measures the constant contribution from the topological entanglement entropy.

\subsection{Entanglement entropy and excited states}

It is a curious property of the toric code that the entanglement properties of excited states are the same as for ground states. This has already been observed in \cite{Nest2014}. Since we have constructed the MERA representation of excited states, we have verified that the entanglement entropy, indeed, only depends on the geometric properties of the subsystem and not on whether the system is in the ground state or not.

As we have seen in section \ref{sec:excited}, representing excited states only changes the value of such qubits which are traced out during the coarse graining procedure but it does not change the structure of the network nor the dimension of the bonds between the various tensors. Hence their sum and thus the entanglement entropy does not change. 

This is remarkable in the light of our observation of the topological entanglement entropy. For general systems, the procedure outlined in \cite{Kitaev2006} is expected to work only for the ground state of a system. The reason that the topological entanglement entropy can also be calculated in excited states is that the toric code is a system with very non-generic properties.

\section{Conclusions}

We have presented an extension of a known MERA for the toric code model which includes representations of excited states. Excitations do not change the structure of the network but only of the value onto which qubits are projected when removed during a step of the renormalization procedure.
From this representation the entanglement properties of all the eigenstates of the toric code can be determined. It is thus easily confirmed that the entanglement properties of all eigenstates are the same, which is a very non-generic property of the toric code model.
Besides, one can discuss important properties like the topological contribution to the entanglement entropy using the MERA network. Although in general the topological entanglement entropy can only be identified in the ground state, due to the entanglement behaviour in the toric code the entanglement entropy can be measured in any state of the system. We have determined that the topological entanglement entropy is, indeed, a constant independent of the geometric size of the patches used to determine it.

The conjectured relation of MERA-networks to the AdS/CFT-correspondence assigns a geometric interpretation to the tensor network. While we have not derived a metric, we can still discuss the geometric interpretation of entanglement properties. Following the prescription of Kitaev and Preskill, we have seen that the geometric contribution of the entanglement entropy corresponds to a defect surface of a specific and constant size when calculating the difference of minimal surfaces which encode the entanglement entropy of appropriately chosen patches. The radial location of the defect surface is located deeper in the bulk and thus more in the IR for larger patches. On the other hand, if we take the patches too small, that is smaller than the correlation length, the measurement of the topological entanglement entropy is in the UV close to the boundary. Here, local effects dominate the geometry, which therefore might not reflect the topological properties at this scale. Of course, in our case, the last consideration does not play a r\^ole since the correlation length vanishes in the toric code. We want to stress again that this geometric picture in MERA is generic, irrespective of the peculiarities of the toric code in other respects.

The toric code is a useful model to determine the holographic description analytically as an approach to MERA as a generalized structure of AdS/CFT. 
Although the model is not generic with respect to the entanglement properties of excited states, it can yet be expected that its general features also apply to low excited states for more generic models. 

\acknowledgments

We gratefully acknowledge discussions with Rom\'an Or\'us and Matteo Rizzi.

\end{document}